\documentclass[]{elsarticle} 
\usepackage[hyphens]{url}

  \journal{Applied Soft Computing} 

\usepackage{lineno} 
\providecommand{\tightlist}{%
  \setlength{\itemsep}{0pt}\setlength{\parskip}{0pt}}

\usepackage{graphicx}
\usepackage{booktabs} 

\usepackage[T1]{fontenc}
\usepackage{lmodern}
\usepackage{amssymb,amsmath}
\usepackage{ifxetex,ifluatex}
\usepackage{fixltx2e} 
\IfFileExists{upquote.sty}{\usepackage{upquote}}{}
\ifnum 0\ifxetex 1\fi\ifluatex 1\fi=0 
  \usepackage[utf8]{inputenc}
\else 
  \usepackage{fontspec}
  \ifxetex
    \usepackage{xltxtra,xunicode}
  \fi
  \defaultfontfeatures{Mapping=tex-text,Scale=MatchLowercase}
  
\fi
\IfFileExists{microtype.sty}{\usepackage{microtype}}{}
\usepackage[top=25mm, left=30mm, right=30mm, bottom=25mm,headsep=10mm, footskip=12mm]{geometry}
\usepackage{longtable}
\ifxetex
  \usepackage[setpagesize=false, 
              unicode=false, 
              xetex]{hyperref}
\else
  \usepackage[unicode=true]{hyperref}
\fi
\hypersetup{breaklinks=true,
            bookmarks=true,
            pdfauthor={},
            pdftitle={Forecasting COVID-19 daily cases using phone call data},
            colorlinks=false,
            urlcolor=blue,
            linkcolor=magenta,
            pdfborder={0 0 0}}
\usepackage{adjustbox,lscape}
\usepackage{float}
\usepackage{booktabs}
\usepackage{longtable}
\usepackage{array}
\usepackage{multirow}
\usepackage{wrapfig}
\usepackage{colortbl}
\usepackage{pdflscape}
\usepackage{tabu}
\usepackage{threeparttable}
\usepackage{threeparttablex}
\usepackage[normalem]{ulem}
\usepackage{makecell}

\begin{document}
\begin{frontmatter}

  \title{Forecasting COVID-19 daily cases using phone call data}
    \author[Cardiff Business School]{Bahman Rostami-Tabar\corref{1}}
   \ead{rostami-tabarb@cardiff.ac.uk} 
    \author[Cardiff School of Computer Science and Informatics]{Juan F. Rendon-Sanchez\corref{1}}
   \ead{rendonsanchezj@cardiff.ac.uk} 
      \address[Cardiff Business School]{Cardiff Business School, 3 Colum Drive, CF10 3EU, Cardiff, UK}
    \address[Cardiff School of Computer Science and Informatics]{Cardiff School of Computer Science and Informatics, Queen's Buildings, 5 The Parade, Roath, CF24 3AA, Cardiff, UK}
      \cortext[1]{Corresponding Author}
  
  \begin{abstract}
  The need to forecast COVID-19 related variables continues to be pressing as the epidemic unfolds. Different efforts have been made, with compartmental models in epidemiology and statistical models such as AutoRegressive Integrated Moving Average (ARIMA), Exponential Smoothing (ETS) or computing intelligence models. These efforts have proved useful in some instances by allowing decision makers to distinguish different scenarios during the emergency, but their accuracy has been disappointing, forecasts ignore uncertainties and less attention is given to local areas. In this study, we propose a simple Multiple Linear Regression model, optimised to use call data to forecast the number of daily confirmed cases. Moreover, we produce a probabilistic forecast that allows decision makers to better deal with risk. Our proposed approach outperforms ARIMA, ETS and a regression model without call data, evaluated by three point forecast error metrics, one prediction interval and two probabilistic forecast accuracy measures. The simplicity, interpretability and reliability of the model, obtained in a careful forecasting exercise, is a meaningful contribution to decision makers at local level who acutely need to organise resources in already strained health services. We hope that this model would serve as a building block of other forecasting efforts that on the one hand would help front-line personal and decision makers at local level, and on the other would facilitate the communication with other modelling efforts being made at the national level to improve the way we tackle this pandemic and other similar future challenges.
  \end{abstract}
   \begin{keyword} COVID-19, time series, forecasting, call centers, regression, exponential smoothing, ARIMA\end{keyword}
 \end{frontmatter}

\hypertarget{introduction}{%
\section{Introduction}\label{introduction}}

Since its discovery at the end of 2019 in Wuhan, China, the spread of coronavirus (SARS-CoV-2) has shaken one country after another putting governments, infrastructure, local and international cooperation to the test as it has claimed the lives of more than a million people worldwide (World Health Organisation 2020) and caused severe disruptions to every-day life and the economy.

The epidemic is evolving with regional and local differences. The way it spreads, as pointed out by Hamzah et al. (2020), is largely influenced by each country's policies. Consequently, in the UK, some of the regional differences might be strongly linked to differences in approaches or policies or their timing to tackle the epidemic in England, Wales, Scotland and Northern Ireland. Additionally, other factors, linked to human direct interaction and exposure, such as life conditions (including income, accommodation, vulnerability), engagement in economic activities with close interaction and mobility patters can further create differences in the evolution of the epidemic at regional and local levels. These complexities are being researched in interdisciplinary areas such as digital epidemiology (Perc et al. 2020; Salathe et al. 2012).

As the epidemic unfolds, decision makers faces the difficulties posed by uneven developments and model limitations to predict its evolution. Increasing numbers of confirmed cases lead to an increase in demand for hospital beds and in the most severe cases also ICU beds. The uneven spread of the infection, manifested in large variations at local level, render important to monitor its spread and attempt to predict the number of new cases. The ability to accurately forecast the number of positive confirmed cases few weeks in advance, would allow local decision makers to put in place all measures required to increase capacity in hospital beds to treat COVID-19 patients before it occurs, instead of reacting to it. This could be achieved, for example, by canceling non-urgent hospital operations and redirect available medical resources to COVID-19 wards. Therefore, an early warning results from an accurate forecasting model would also allow other services such as mortuary services to plan ahead for extra capacity in body storage.

Frustration with regards to accuracy when forecasting COVID-19 has been summarised by Ioannidis, Cripps, and Tanner (2020). A myriad of reasons have been given, including poor data, lack of incorporation of epidemiological features and lack of transparency. We have also observed some limitations when reviewing the literature on forecasting different COVID-19 variables which will be summarised as follows: i) the forecast accuracy of models is not systematically reported; ii) there are big differences between predictions and what happened in reality, forecast errors greatly vary; iii) in some studies it is unclear how forecasts are generated and evaluated, e.g.~whether a rolling window is used, ,what si the forecast horizon; and iv) most of the studies focus on point forecasts and ignore the uncertainty associated with them, which may lead to incorrect policies.

Despite these efforts and their limitations, the need to have appropriate tools to aid decision making continues to be pressing and sensible contributions within forecasting are sought after. On the one hand, decision makers at national and local levels need support for general strategy, but on the other hand, clear orientation at ground level, when touching more closely with the reality, is is crucial to adjust operations and resources. The communication between these different levels is also acutely needed and therefore any experience with different forecasting and other decision tools at local level with a clear assessment of their accuracy will facilitate the integration of models and communication as the epidemic continues to unfold.

As part of such effort, this paper concentrates on the task of foresting COVID-19 cases at a local level by using call data. It is a contribution towards enlarging the set of tools and accuracy assessment, to aid front line health personnel and managers facing the crisis in taking decisions on resources and operations. This study has also been motivated by discussions with practitioners in a country council in England highlighting the importance of having a model to forecast the number of cases at the local level.

In this paper, our objectives are fourfold: (i) we propose a model to accurately forecast daily COVID-19 confirmed cases; (ii) we examine the use of call data and provide evidence of their usefulness in forecasting daily confirmed cases; (iii) we provide probabilistic forecasts that quantify uncertainties in future confirmed cases; (iv) we benchmark the accuracy of our model against two time series techniques including 1) AutoRegressive Integrated Moving Average, ARIMA (Box et al. 2015), 2) exponential smoothing state space model, ETS (Durbin and Koopman 2012) and a regression model without considering call data.

The rest of the paper is organised as following: Section \ref{lit} provides a brief overview of the efforts to forecast COVID-19, Section \ref{design} starts with analysing data followed by introducing the models used in this study and the forecast performance evaluation scheme and metrics. Section \ref{result} presents the result of the study followed by concluding remarks in section \ref{conclusion}.

\hypertarget{lit}{%
\section{Research background: forecasting COVID-19}\label{lit}}

COVID-19 has attracted a lot of attention of researchers to forecasting and may studies have analysed various related variables. Some studies focus on the global and national level; others focus on regional level and few consider the local level. The following is an overview of approaches use to forecast different variables of COVID-19 to date.

Petropoulos and Makridakis (2020) forecast the cumulative number of daily confirmed cases globally with exponential smoothing models. Since this early effort, the limitations of the models were clear and acknowledged by the authors and attributed to strong government interventions and accuracy of data. Benvenuto et al. (2020) predict the epidemiological trend of incidence and prevalence of COVID-19 worldwide using ARIMA models. Feroze (2020) use Bayesian Structural Time Series Models Brodersen et al. (2015) to forecast the daily total number of positive cases in USA, Brazil, Russia, India and UK for the next 30 days. Authors report a better performance from their models when compared to ARIMA models.

Yousaf et al. (2020) forecast daily confirmed cases with ARIMA models at country level in Pakistan. Of the same nature is the research by Moftakhar, Mozhgan, and Safe (2020) with data from Iran, in which observed new cases are used to predict the number of patients. Artificial Neural Networks (ANN) and Auto-Regressive Integrated Moving Average (ARIMA) are used, with the later showing better accuracy. ARIMA models are also used by Gupta and Pal (2020) to forecast infection cases in India for the next 30 days. Harun et al. (2020) forecast the number of COVID-19 cases in Germany, United Kingdom, France, Italy, Russia, Canada, Japan, and Turkey by using linear and non-linear regression models, ARIMA and exponential smoothing methods. Different best performing forecasting models are identified for different countries.

Tomar and Gupta (2020) use a long short-term memory (LSTM) recurrent neural network to forecast daily and total positive cases, total recovered and total deceased in India. A simple curb fitting with an exponential model is used to assess the effect of a lock-down and social distancing. LSTM networks were also used with data from Canada by Chimmula and Zhang (2020).

Perc et al. (2020) forecast the number of daily confirmed cases in United States, Slovenia, Iran and Germany with an iterative model that uses the average growth of cases and takes into account the recovered and deceased. Having devised the model around growth and targeted its used to the short term, renders it helpful even in the absence of a systematic accuracy study.

Logistic models of the SEIRD type, which use groups of population (Susceptible, Exposed, Infected, Removed and Dead), and are based on the solution of a system of differential equations, have been used abundantly, with different variants, to describe and predict the evolution of the epidemic globally (Hamzah et al. 2020) and in regions and countries (Martelloni and Martelloni 2020; Fanelli and Piazza 2020; Mohd and Sulayman 2020; Nabi 2020). Martelloni and Martelloni (2020) apply a SIRD model to Italy and adapt it to account for the introduction of lock-downs by governments, the influence of asymptomatic and the absence of a criteria to define the susceptible group. Fanelli and Piazza (2020) use a SIRD model to explore the temporal dynamics of the epidemic in China, Italy and France. Sarkar, Khajanchi, and Nieto (2020) adapt the SIRD model to account for asymptomatic, isolated infected and quarantined susceptible. Mohd and Sulayman (2020) propose a modified logistic model to account for reinfections, false detection problems and scarcity of medical equipment in the context of a developing country and found that in such circumstances there can be unstable scenarios where decisions based on the reproduction number \(R_0\) should be made with caution. In some studies (Martelloni and Martelloni 2020; Fanelli and Piazza 2020) there are hints of universal characteristics, whereas in Mohd and Sulayman (2020) the particular circumstances of a developing country suggests that behaviours of the epidemic might have important country differences.

Sardar et al. (2020) investigat several models covering the main two categories, mathematical (SIRD) and statistical, to study the effect of lock-down in India: a SIRD-type logistic model, expanded further than Sarkar, Khajanchi, and Nieto (2020) that accommodates for lock-down; an Auto-regressive Integrated Moving Average (ARIMA) model; an exponential smoothing model with ARMA errors, trend and seasonal components (TBATS); a hybrid statistical model based on the combination of ARIMA and TBATS and a weighted combination of the SIRD model and the best statistical one. Daily notified cases from five states in India and the overall country are used to fit the models, which helped assessing different scenarios determined by the effective reproduction rate number.

Sophisticated spatio-temporal models have been explored by The Gleam Project (2020), Balcan et al. (2009) and Balcan et al. (2010) to project infections, deaths and ICU and hospital beds needed in the US. This and related work, such as Liu et al. (2020), indicate the need of complex methods for some settings.

At regional level, Anastassopoulou et al. (2020) propose a methodology to estimate epidemiological parameters and applied a SIRD model to forecast the number of infections, recovered and deceased in the region of Hubei, China. After a coarse estimation of parameters, a non-linear optimisation routine is used to obtain refined estimates. The official figures from the region fell inside the upper and lower bounds, albeit very wide, that had been produced by authors. Despite the inaccuracy, the lessons learned in how to calibrate this type of models are very valuable.

Massonnaud, Roux, and Crépey (2020) adapt a SEIR model to formulate scenarios for French metropolitan regions based on different reproduction rates. The daily number of COVID-19 cases, hospitalisations and deaths, the needs in ICU beds per region and the reaching date of ICU capacity limits were estimated. Hospital catchment areas are used and then aggregated by French region.

Weissman et al. (2020) implement a SIR model to plan scenarios (defined by doubling times of 2, 6 and 10 days) in 3 hospitals in greater Philadelphia region, US. Three variables are predicted: hospital capacity, patients requiring ICU beds and patients requiring ventilators. The accuracy of the predictions is not reported, but the usefulness of the scenarios in the planning ahead is recognised. This assessment and the construction of models is facilitated by a close collaboration between operational, medical and data personnel.

When considering these forecasting efforts, it is clear that accuracy has been disappointing in many cases. Majority of these studies focus only on point forecasts and ignore the uncertainty that may lead to incorrect policies. in some cases, the forecast accuracy has not been reported or it is not clear how it has been evaluated. It is also evident the effort in expanding the SIRD models in an attempt to adapt them to different circumstances and government interventions. However, their limitations might not be overcome unless connected with other approaches, such as spatio-temporal complex models like GLEAM. Nonetheless, despite its crudeness, SIRD models are helpful at regional level in facilitating the visualisation of different scenarios to guide planning and decision making (Fanelli and Piazza 2020; Massonnaud, Roux, and Crépey 2020; Weissman et al. 2020).

\hypertarget{design}{%
\section{Experimental design}\label{design}}

\hypertarget{data}{%
\subsection{Data}\label{data}}

Data used in this paper comprised the number of daily COVID-19 confirmed cases and the number of daily phone calls received at the National Health Service 111 (NHS 111) at one of the largest Non-metropolitan country council in the East Midlands region of England between 18 March 2020 and 19 September 2020. Data is publicly available and extracted from the Public Health UK\footnote{https://coronavirus.data.gov.uk/} and NHS Digital\footnote{https://digital.nhs.uk/coronavirus}, respectively.

\begin{figure}[H]

{\centering \includegraphics[width=0.7\linewidth,]{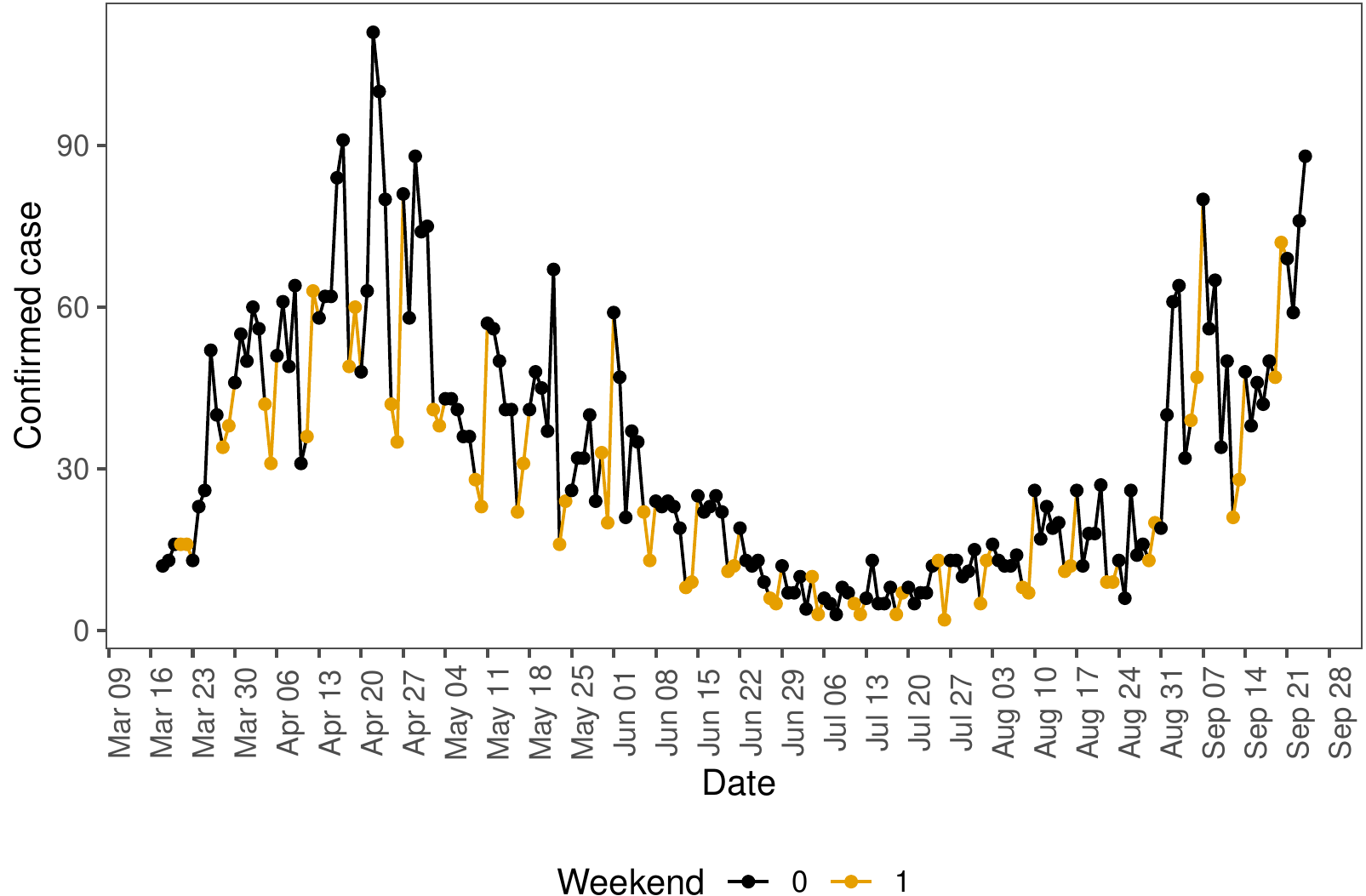} 

}

\caption{Time series of daily confirmed cases}\label{fig:weekend}
\end{figure}

\begin{figure}[H]

{\centering \includegraphics[width=0.7\linewidth,]{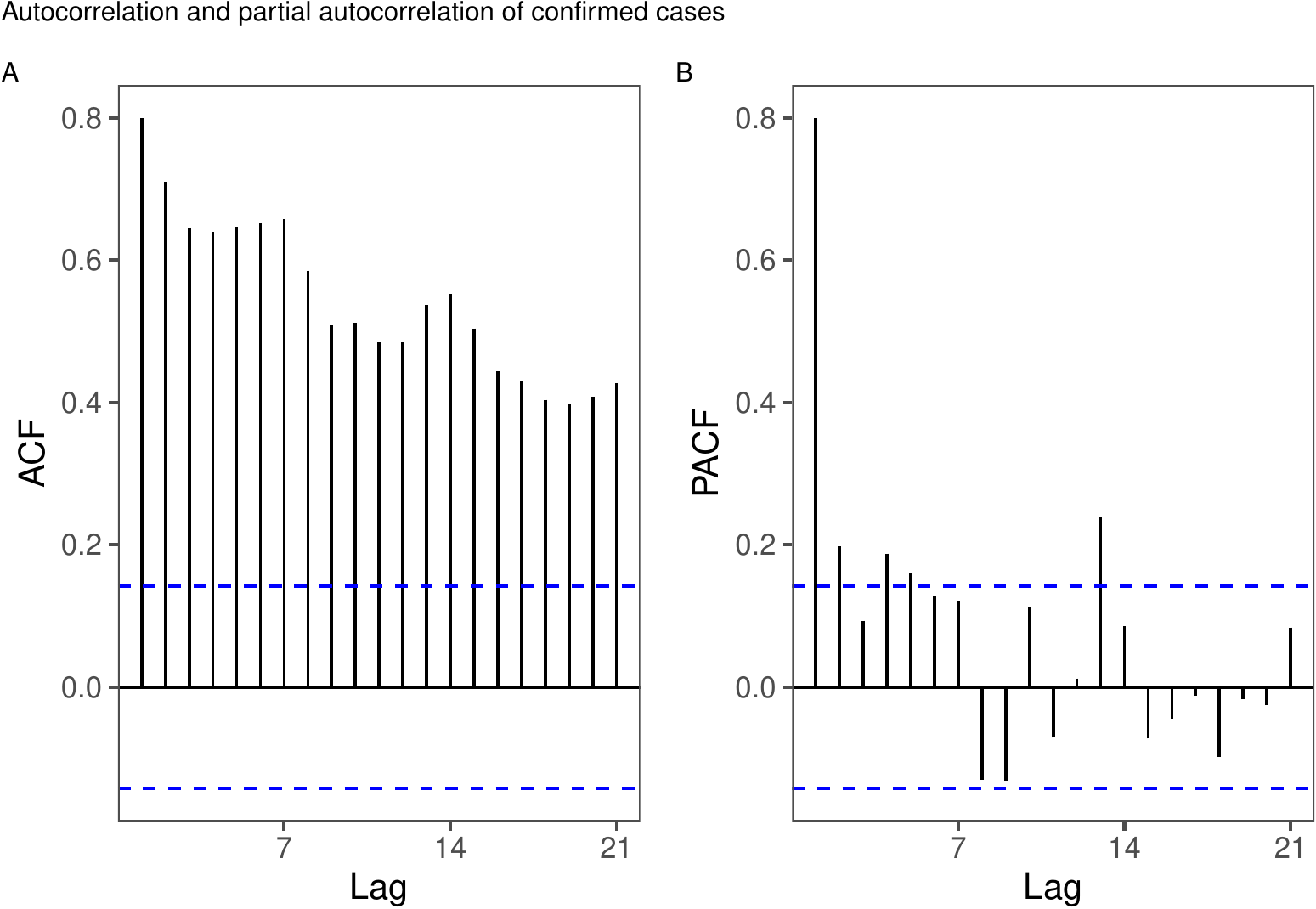} 

}

\caption{Autocorrelation and Partial Autocorrelation of the time series of confirmed cases}\label{fig:acf}
\end{figure}

Figure \ref{fig:weekend} illustrates the time plot of the confirmed cases. Although the time series is noisy, some systematic patterns are visible. The time plot shows a trend and the number of positive cases in the weekend are lower than the working days. We also analyse the Autocorrelation (ACF) and Partial Autocoirrelation (PACF) Functions to investigate the link between the number of confirmed cases and its lagged values. Figure \ref{fig:acf} shows the ACF (Figure \ref{fig:acf}-A) and PACF (Figure \ref{fig:acf}-B) of the time series. The ACF plot highlights a trend because by increasing the lag, the value of ACF decreases exponentially. Additionally, high positive spikes in \(lag 7 ,14 and 21\) shows the presence of seasonality. Moreover, PACF reveals that there are some positive lags that need to be considered in the forecasting model such as \(lag 1\), \(lag 4\), \(lag 5\), and \(lag 13\).

\begin{figure}[H]

{\centering \includegraphics[width=0.7\linewidth,]{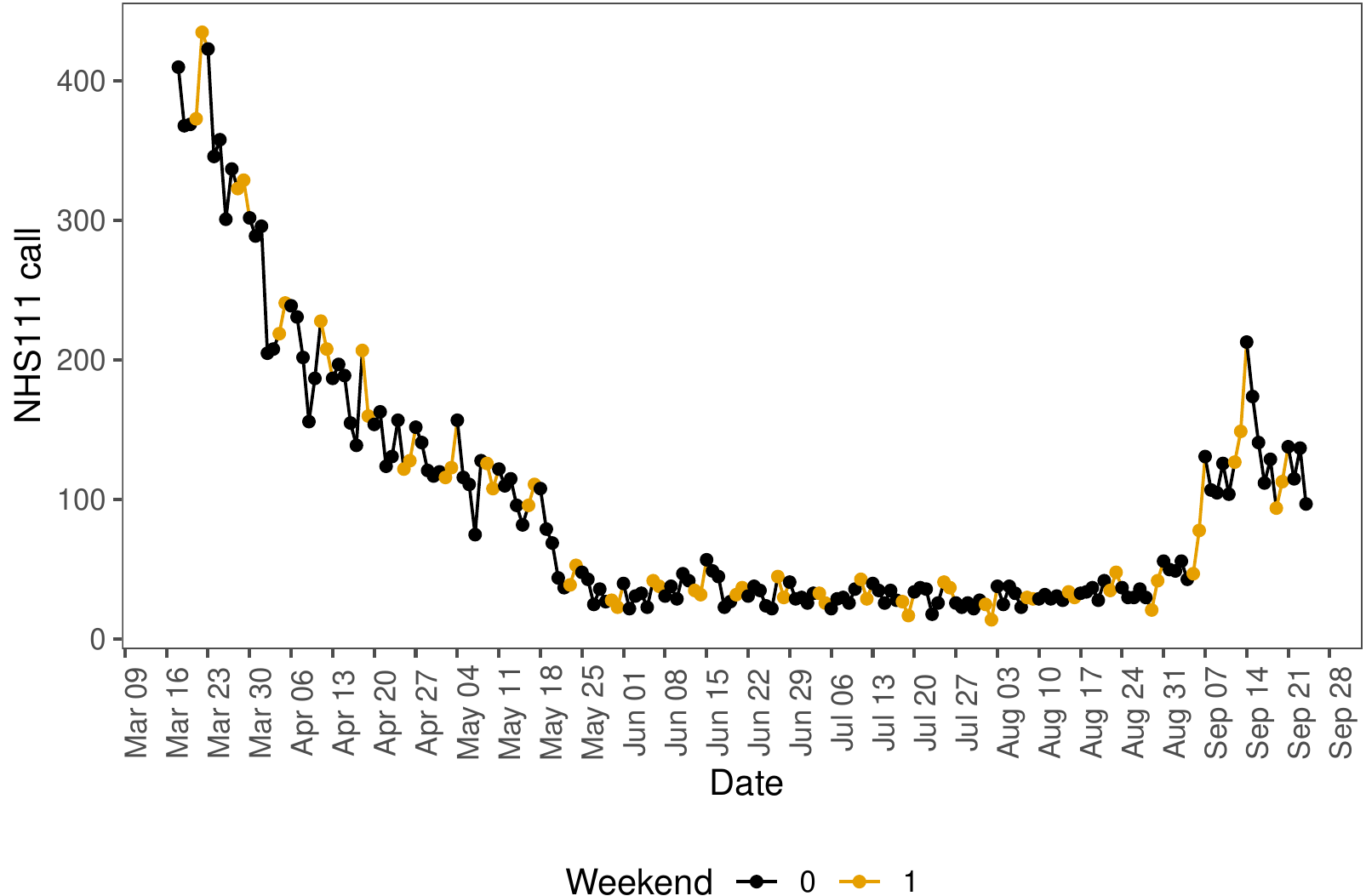} 

}

\caption{Time series of daily calls.}\label{fig:NHS111}
\end{figure}

Figure \ref{fig:NHS111} presents the time plot of the NHS 111 calls highlighting the presence of a trend. The data is less noisy and there is no seasonal pattern. NHS 111 call data can be used as a predictor of the number of daily confirmed cases, therefore, we need to produce its forecast.

In the relationship between the number of confirmed cases and the number of NHS 111 calls on day \(t\), the former may be related to past lags of the later. The sample Cross-Correlation Function (CCF) can be used to identify lags of the number of NHS 111 calls series that might be useful predictors of the number of confirmed cases. For instance, consider \(lag = -24\), the CCF value will give the correlation between the number of NHS 111 calls on day \(t-24\) and the number of confirmed cases on day \(t\).

\begin{figure}[H]

{\centering \includegraphics[width=0.7\linewidth,]{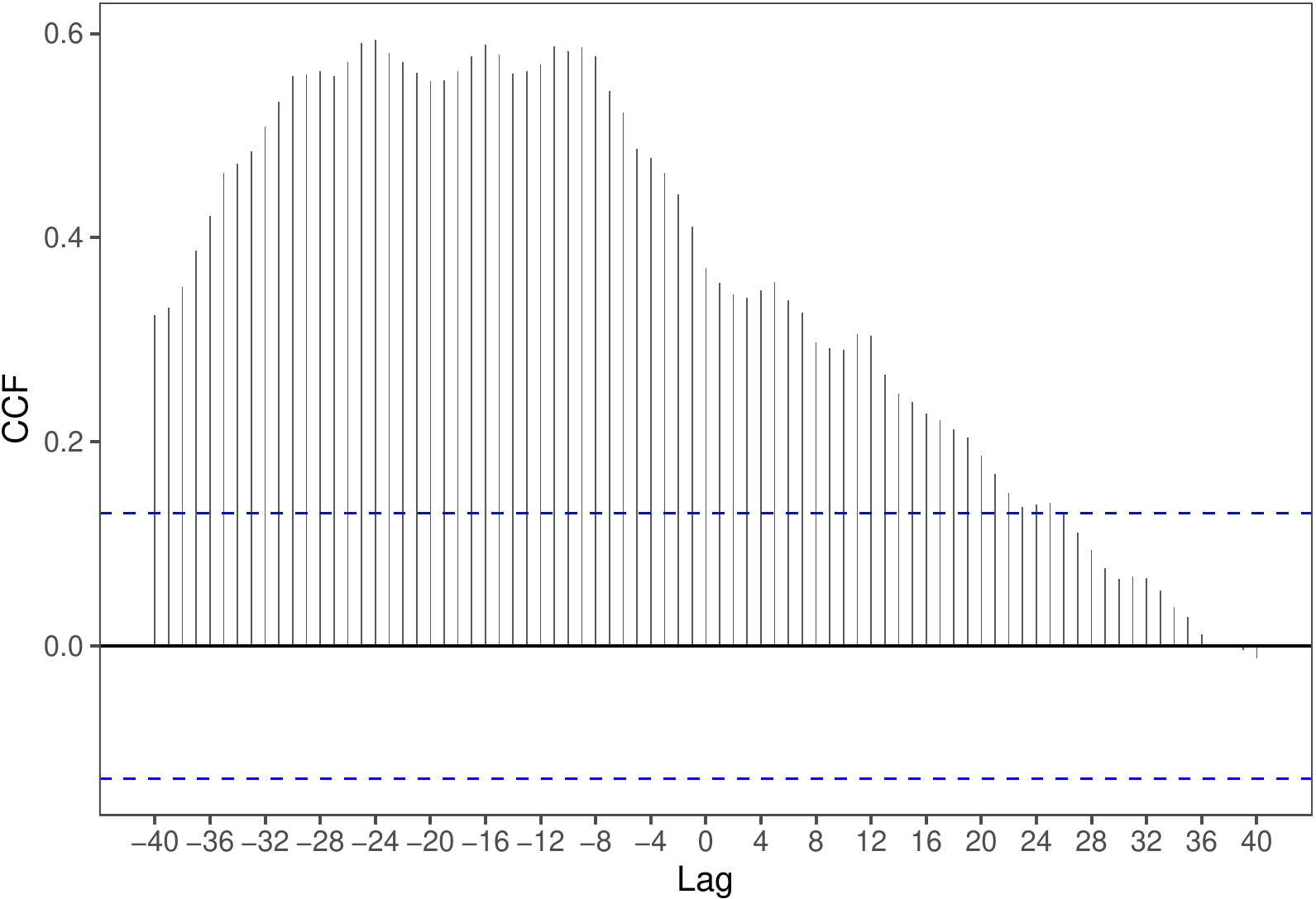} 

}

\caption{Cross-correlation of the past values of NHS 111 calls and the number of confirmed cases.}\label{fig:ccf}
\end{figure}

Figure \ref{fig:ccf} presents the linear relationship between confirmed cases and NHS 111 lags. It seems that confirmed cases are relatively high almost \(7-30\) days after high NHS 111 calls (i.e., significant correlation at lags of \(7\) to \(30\)). There are a lot of models that we could try based on the CCF, ACF, PACF and the presence of trend and the effect of weekends for these data. In the next section, we propose a Multiple Linear Regression model in which the number of confirmed cases, is a linear combination of (past) lags of the number of NHS 111 calls.

\hypertarget{pmodel}{%
\subsection{Proposed model, Multiple Linear Regression with call data (MLR\_T)}\label{pmodel}}

In this section, we model the number of confirmed cases as potentially a function of past lags of confirmed cases, current and past lags of the number of NHS 111 calls, the weekend and the trend. These components allow possible interpretations as described in section @\ref(sec:data).

We denote \(Y_t\) as the number of positive confirmed cases on day \(t\). The predictors are presented using the following notations:

\begin{itemize}
\tightlist
\item
  \(NHS 111_{t}\): for the number of NHS 111 calls on day \(t\).
\item
  \(trend_{t}\): for the local trend on day \(t\).
\item
  \(Y_{t-k}\): for the number of positive confirmed cases on day \(t-k\).
\item
  \(NHS 111_{t-k-4}\): for the NHS 111 calls on day \(t-k-4\).
\item
  \(Weekend_{t}\): for the weekend dummy; \(Weekend_{t}\) is 1 if \(t\) falls on a Weekend and 0 otherwise.
\end{itemize}

\begin{equation}
Y_t = \beta_0 + \beta_1 \text{NHS 111}_t+ \text{trend}_{t}+
\underbrace{\sum_{k=1}^{21} \beta_{k+1}
\text{Y}_{t-k}
}_{\text{AutoRegressive effect}}+ \underbrace{\sum_{k=1}^{26} \beta_{22+k} \text{NHS 111}_{t-k-4}}_{\text{NHS 111 lag effect}} + Weekend_t + \varepsilon_t \label{eq:pm}
\end{equation}

The mathematical representation of the proposed model is shown in the Equation \eqref{eq:pm}. We should note that, to ensure achieving strictly positive values for the number of confirmed cases, we have used the \texttt{log()} transformation for the response variable, confirmed cases, and its lagged values as predictors. Specified transformation is automatically back-transformed to produce forecasts.

Not all potential predictors are included in the final forecasting model. After building the model, we need to select the predictors that are useful in producing accurate forecasts. To that end, we have conducted a stepwise regression and select predictors that improve the Out-of-Sample forecast accuracy based on RMSE. Due to the amount of uncertainty in the number of positive confirmed cases, we produce the estimated probability distribution of the number of positive confirmed cases in future time periods in addition to the point forecast. This allows us to provide more information about the forecasts that could be fruitfully used by decision makers.

The proposed model uses the NHS 111 call, its lags and the lagged values of confirmed cases as predictors, therefore, in order to produce ex-ante forecasts (Hyndman and Athanasopoulos 2020), we need to generate forecasts of these predictors.

To forecast NHS 111 calls, we examined ARIMA, ETS and their combination. We use forecasts generated by ARIMA for NHS 111 calls as it results in more forecast accuracy. For the lagged values of confirmed cases, we examines Naive (Hyndman and Athanasopoulos 2020), ARIMA, ETS and their combinations. Forecasts generated by ETS result in more accurate forecasts overall, however, forecasts from \texttt{Naive} method were accurate for shorter horizons. Therefore, we replace the forecasts generated from ETS by those generated from the \texttt{Naive} method when \(h < 5\). We have compared various horizons, \(h\) and this yields in a more accurate forecast of confirmed cases in the Out-of-sample.

\begin{figure}[H]

{\centering \includegraphics[width=0.9\linewidth,]{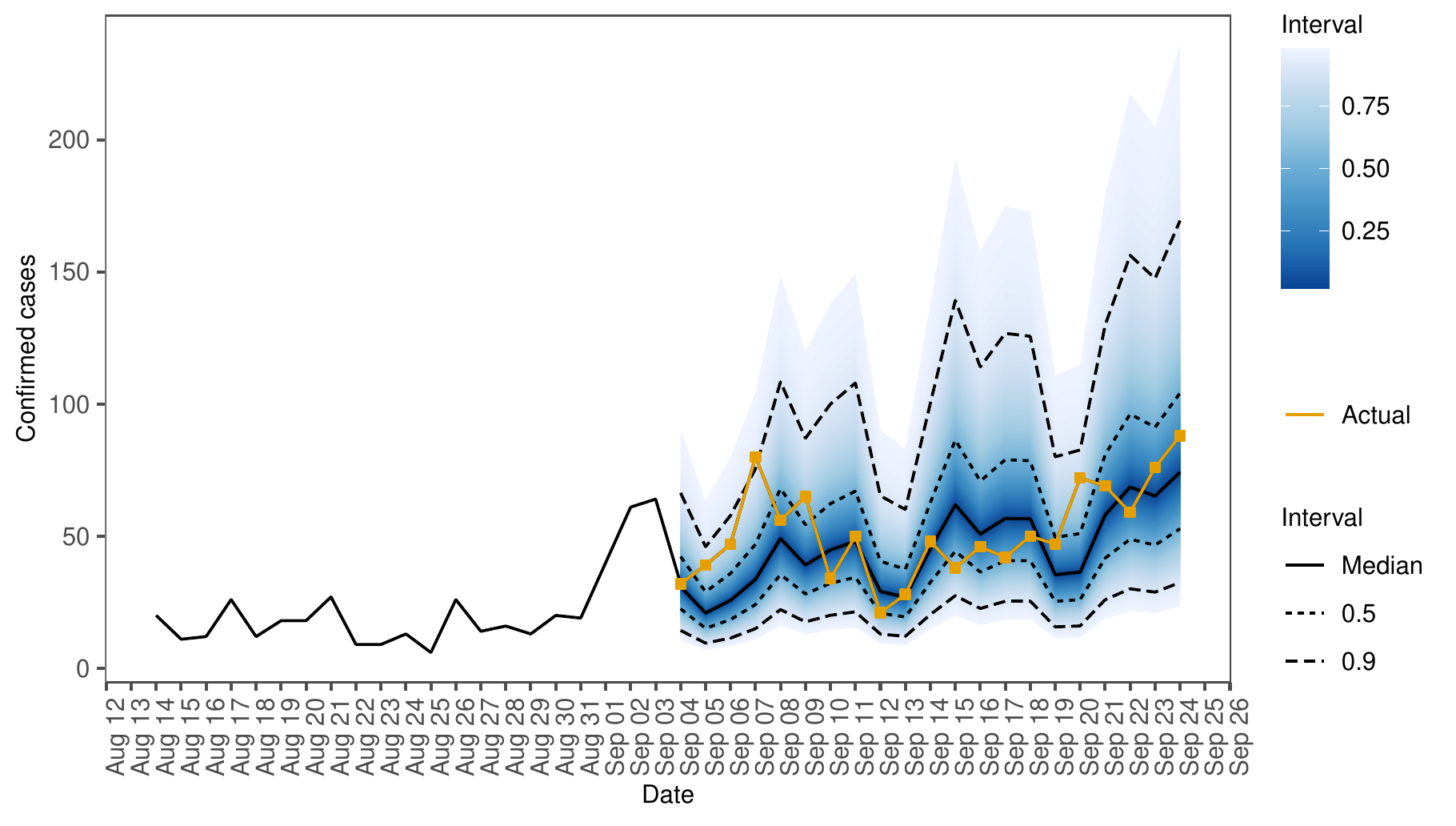} 

}

\caption{Sample forecast produced by the proposed model.}\label{fig:samplefcst}
\end{figure}

Figure \ref{fig:samplefcst} illustrates an example of the probabilistic forecast generated by the proposed model.

\hypertarget{benchmarks}{%
\subsection{Benchmarks}\label{benchmarks}}

We use three different benchmark methods to compare their forecast accuracy with our proposed model. We consider Exponential smoothing and ARIMA models as the two most widely used approaches to time series forecasting, in addition to a regression approach without using call data. These benchmark methods are used to show whether there is any adding value in using the proposed model or not.

\hypertarget{proposed-model-without-call-data-mlr_w}{%
\subsubsection{Proposed model without call data (MLR\_W)}\label{proposed-model-without-call-data-mlr_w}}

In order to show the impact of incorporating the call data in the modeling and its effect on forecast accuracy improvement, we use the the same model proposed in section \ref{pmodel} without including NHS 111 calls and its lagged variables.

\hypertarget{exponential-smoothing-state-space-model-ets}{%
\subsubsection{Exponential smoothing state space model (ETS)}\label{exponential-smoothing-state-space-model-ets}}

The second benchmark is the automatic exponential smoothing (Hyndman and Athanasopoulos 2020) model that is incorporated using corresponding implementation of the fable package in R. We use the \texttt{ETS()} function in the fable package (O'Hara-Wild et al. 2020) to generate daily forecast of the confirmed cases.

\hypertarget{autoregressive-integrated-moving-average-arima}{%
\subsubsection{AutoRegressive Integrated Moving Average (ARIMA)}\label{autoregressive-integrated-moving-average-arima}}

The third benchmark model is an automatic ARIMA) model (Hyndman and Athanasopoulos 2020). The orders of the model and parameters are selected by minimising the AIC (Akaike information criterion). We use the implementation of automatic ARIMA in \texttt{fable} package; \texttt{ARIMA()} function is used to generate the forecasts of daily confirmed cases (O'Hara-Wild et al. 2020).

\hypertarget{accuracy}{%
\subsection{Forecast performance evaluation schemes and metrics}\label{accuracy}}

We first split the dataset into training (\(70\%\)) and test (\(30\%\)) sets. We apply the forecasting models on the training set and evaluate the forecast performance of the proposed MLR approach and benchmarks on the test set. Evaluation is conducted using a rolling origin forecasting study with re-estimation. Forecasting horizon is chosen to be \(h=21\) which corresponds to three weeks that allows decision makers to use forecast to inform decisions on resource allocation and planning in the health service. We have considered six different forecast accuracy metrics: i) three point forecast error metrics, ii) one prediction interval accuracy measure, and iii) two probabilistic forecast accuracy metrics.

\hypertarget{point-forecast-accuracy-measures}{%
\subsubsection{Point forecast accuracy measures}\label{point-forecast-accuracy-measures}}

The point forecasting metrics are the ME (Mean Error), MAE (mean absolute error) and RMSE (root mean square error) for point forecasts (Hyndman and Koehler 2006). These measures help us to evaluate the performance of models from different perspectives.
ME provides the overall direction of the error. It reveals whether the produced forecasts on average are too high or too low. MAE is strictly appropriate for the median, i.e.~if the MAE of a model is smaller than others, it produces forecasts closer to the median of the data than the others. Finally, RMSE is minimised for the optimal mean forecast. That means a model with the lower value of RMSE produces more accurate mean values comparing to others. We evaluate the ME, MAE and RMSE for each model and each forecasting step and separately. We also report the overall performance of each model across all horizons. A model with an error measure closer to zero is better.

\hypertarget{prediction-interval-accuracy-measures}{%
\subsubsection{Prediction interval accuracy measures}\label{prediction-interval-accuracy-measures}}

An appropriate prediction interval accuracy measure should account for both the coverage and the width of the prediction interval. Winkler (1972) proposed a measure to calculate the accuracy of the forecast prediction interval for any given model.

If \(y_t\) is the observation at time \(t\) and \([l_{t}, u_{t}]\) represents the \(100(1 - \alpha)\%)\) prediction interval at time \(t\), then the score for each period is:

\begin{equation}
  W(l_t,u_t,y_t) =
    \begin{cases}
      u_t-l_t & \text{if $l_t < y_t <u_t$}\\
      (u_t-l_t)+ \frac{2}{\alpha}(l_t-y_t) & \text{if $y_t < l_t$}\\
      (u_t-l_t) + \frac{2}{\alpha}(y_t-u_t) & \text{if $y_t > u_t$}
    \end{cases}
\end{equation}

The Winkler score is the average of scores across all periods. We prefer a model that has a narrow prediction interval with high coverage, therefore a model with smaller score is better. In this study, Winkler score is calculated based on \(95\%\) prediction intervals.

\hypertarget{probabilistic-accuracy-measures}{%
\subsubsection{Probabilistic accuracy measures}\label{probabilistic-accuracy-measures}}

In addition to the point and prediction interval accuracy measures, we also report two probabilistic forecasting measures: i) Percentile score and , ii) Continuous Ranked Probability Score (CRPS).

The percentile score is a strictly appropriate evaluation criterion for quantiles. The evaluation is perfomed on a dense probability grid for all percentiles (\(1\%,\ldots,99\%\)) (Hong et al. 2016). For each time period \(t\), we obtain the quantile, \(q_{i,t}\), where \(i = 1, 2, \ldots , 99\). Then, the percentile score is given by the pinball loss function at each period t:

\begin{equation}
  L(q_{i,t}, y_t) =
    \begin{cases}
    (1 - i/100)(q_{i,t} - y_t) & y_t < q_{i,t} \\ (i/100)(y_t - q_{i,t}) & y_t \geq q_{i,t}.
    \end{cases}
\end{equation}

The score can then be calculated for each forecast horizon and across all percentiles. We also report the overall percentile score across all horizons. If the observations does not deviate from the forecast distribution, then the average score is small. A model with smaller percentile score is better. CRPS summaries the quality of a continuous probability forecast with a single score by measuring the distance between the forecast and the observed cumulative Distribution Function. It reveals how closely the CDF of the forecast matches that of the corresponding observations.

We should note that in order to report the result of the performance evaluation using time series cross validation, we have averaged the score of each measure computed for different rolling origins.

\hypertarget{result}{%
\section{Result and discussion}\label{result}}

In this section, we present the forecast accuracy of our MLR model and its benchmarks for point forecast, prediction interval and probabilistic forecasts using Out-of-sample data based on time series cross validation.

Table \ref{tab:point} shows the overall performance of models for all type of point, prediction interval and probabilistic accuracy measures averaged across all forecast horizons. The proposed MLR clearly outperforms benchmarks for all forecast accuracy measures.

We observe that all models on average are over estimating the number of daily confirmed cases. The proposed model has lower bias comparing to ETS and ARIMA. Moreover, if the decision maker is interested in getting the mean or the median number of daily cases, the proposed approach provide better accuracy. We also observe that it has the lowest Winkler score, which means on average it has better coverage and narrower intervals than others. Most importantly, our model provides more accurate probability forecast distribution that others, which means it can inform more accurately about uncertainty.

\begin{figure}[H]

{\centering \includegraphics{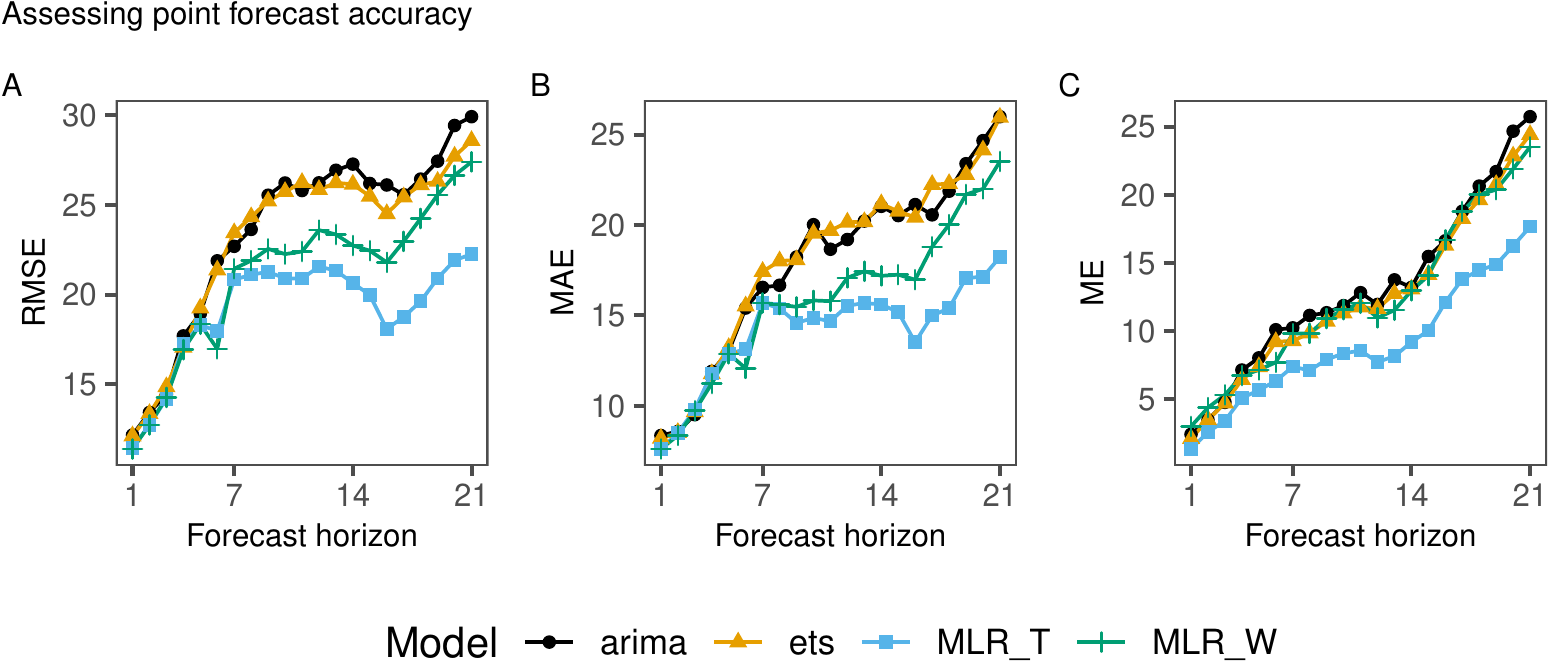} 

}

\caption{Point forecast accuracy of models for each forecast horizon}\label{fig:pointh}
\end{figure}

In addition to the overall performance, we have also analysed the result of each forecast horizon. Figure \ref{fig:pointh} illustrates the point forecast accuracy metrics for each forecast horizon. While for shorter horizons, the performance of models is very similar, there is a substantial gain in longer horizons using the proposed approach. For longer forecasting horizons, the proposed model retains high forecasting accuracy compared to other methods. Clearly, the proposed model not only remains accurate across the forecasting horizon, but is also more robust than other methods against increasing horizons.

\begin{figure}[H]

{\centering \includegraphics{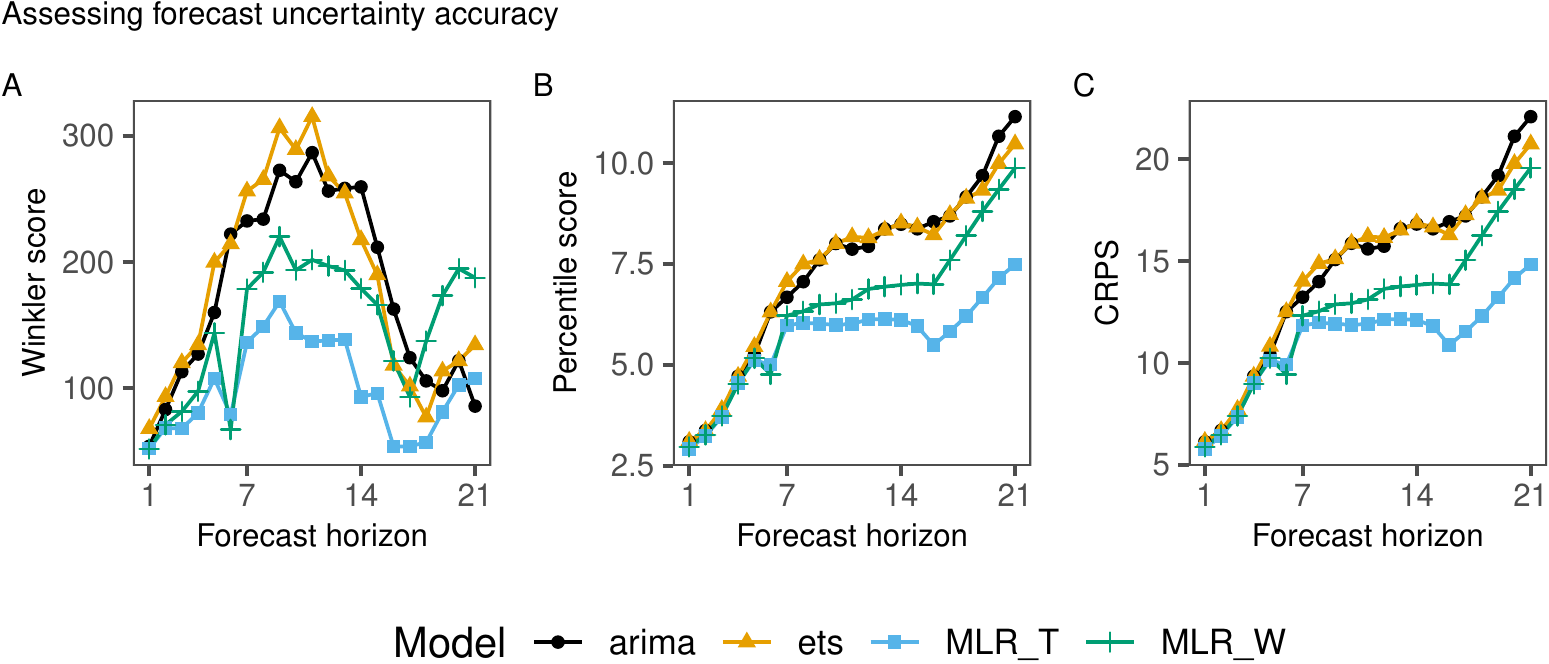} 

}

\caption{Forecast uncertainty of models for each forecast horizon}\label{fig:uncertainty}
\end{figure}

Figure \ref{fig:uncertainty} presents measures of the uncertainty of the models used in this study. Figure (\ref{fig:uncertainty}-A) shows the Winkler score for the \(95\%\) prediction intervals for each horizon. The proposed model provides more accurate prediction interval forecasts for most horizons with an exception of very high horizon in \(h= 19, 20, 21\), where the ARIMA model produces better results.

In evaluating the probabilistic forecast accuracy of our model and its benchmark, we focus on the distributional characteristics of the generated forecasts evaluated by the corresponding percentile score and CRPS. Figure (\ref{fig:uncertainty}-B and C) presents the corresponding percentile score and CRPS aggregated for each forecasting horizon. The results show that the proposed model captures the density structure for most of the horizons better than the benchmark models. The difference is substantial for forecast horizons longer than 7 days.

Our results show that the proposed model not only outperforms benchmarks in terms of generating the correct value of future confirmed cases, but is also the most accurate when providing consistent information about uncertainty around confirmed cases. The model is very simple and can be understood by any manager, it also provides the entire probability distribution for confirmed cases for a given day at the local level. Our model offers the ability to capture a range of possibilities regarding confirmed cases for any given day that is not contained in the point forecasts. This is very important for decision makers and planners in healthcare, as it helps them to assess the risk and make better decisions in planning. Further research attempts are required on how to use probabilistic forecasts and interpret output to inform decisions in the healthcare.

We should note that, if the focus of the decision makes is more on the short horizons ( \(h < 7\)), then it is possible to create a MLR based model using the principles discussed in this paper that outperforms benchmarks.

\begin{table}

\caption{\label{tab:point}Forecast performance evaluation}
\centering
\begin{tabular}[t]{>{}lrrrrrr}
\toprule
\multicolumn{1}{c}{ } & \multicolumn{6}{c}{Accuracy measture} \\
\cmidrule(l{3pt}r{3pt}){2-7}
Model & ME & RMSE & MAE & Winkler & Percentile & CRPS\\
\midrule
arima & 13.14 & 24.06 & 17.88 & 177.64 & 7.37 & 14.60\\
ets & 12.38 & 23.59 & 18.08 & 183.62 & 7.35 & 14.58\\
\textbf{MLR\_T} & \textbf{8.96} & \textbf{19.37} & \textbf{14.16} & \textbf{100.53} & \textbf{5.60} & \textbf{11.10}\\
MLR\_W & 12.35 & 21.47 & 15.82 & 149.51 & 6.43 & 12.76\\
\bottomrule
\end{tabular}
\end{table}

\hypertarget{conclusion}{%
\section{Conclusion}\label{conclusion}}

An accurate forecast of the daily number of positive confirmed cases is crucial for resource planning. Many models have been proposed in the literature to forecast confirmed cases ranging from time series to epidemic forecasting models.

This study has been motivated by a forecasting problem faced by a local country council in England that highlights the lack of a proper method to forecast the number of positive confirmed cases at the local level. We have created a simple and interpretable model that can be used to forecast positive confirmed cases at the local level. The proposed MLR model exploits the relationship between the confirmed case and the call data (NHS 111 calls), in addition to other patterns such as trend, the effect of weekends and Autoregressive lags of confirmed cases.

We compare the performance of the model with ETS, ARIMA and a MLR model without call data. These models are applied to the number of confirmed cases for a country council in England; our analysis showed that the proposed model can provide reliable forecasts. In addition to generating traditional point forecasts, we also provide the probabilistic forecasts. Given the uncertainty involved in forecasting the number of COVID-19 confirmed cases, this is extremely important for decision makers as it highlights the uncertainty around a single value forecast. In evaluating the performance of generated forecasts, we have used three point forecast error measures, one prediction interval accuracy measure and two probabilistic forecasting accuracy measures.

We propose a simple, interpretable and reliable forecasting model that outperforms its benchmarks in generating point forecasts, prediction intervals and probabilistic forecasts for daily number of COVID-19 confirmed cases using an empirical study analysing daily historical confirmed cases and call data. Additionally, in most of the studies related to forecasting COVID-19, forecast uncertainty has not been reported , especially the probability forecast distribution. It provides more information about uncertainty of future confirmed cases, which can be used for risk management in allocating resources. We also provide evidence that incorporating the call data into the forecasting model can improve the forecast accuracy in both point forecasts and probabilistic one.

We have also examined forecasting using the ensemble of methods used in the paper. We observed that while this may improve the point forecast accuracy, but it deteriorates the prediction interval and probabilistic accuracy, so any ensemble method should be used with caution. Although the proposed model has been optimised for a forecast horizon of 21 days, we noticed it is possible to create a new one by focusing on shorter horizons (\(<7\) days), using a similar principle, that is more accurate for for such horizons.

We have also used the proposed model to forecast the number of confirmed cases at the national level, England. The overall performance is very similar to the local level, the proposed model outperforms others. However, we have not provided the analysis here due to the space limit and the focus of the study at the local level, but it can be provided on request.

A more complete assessment of the model could include in the future compartmental benchmarks. However, with the knowledge gathered in this forecasting task, we could propose that the positive performance of the model presented here lies in distancing from the complexity of SEIRD compartmental models while at the same time exploiting a reliable source of information, phone calls, that is ultimately needed when departing from theoretically informed models. SEIRD models have to be expanded to account for lock-downs, and other interventions and eventualities unfolding at the same time than the pandemic, and also have to be carefully initialised. Cutting through this complexity with a well informed simple model is a potentially meaning contribution for decision makers. Further improvements of the proposed model can be achieved by incorporating additional information, such as local regulations or theoretical elements of compartmental models. Moreover, one may try to use exponential smoothing and ARIMA models with call data as exogenous variables.

\hypertarget{acknowledgments}{%
\section*{Acknowledgments}\label{acknowledgments}}
\addcontentsline{toc}{section}{Acknowledgments}

We thank Mr.~Sergio Pappalettera, Principal Analyst -- Population Health Management (PHM), NHS Nottingham and Nottinghamshire CCG for the useful discussions, highlighting the importance of creating a model to forecast new cases at the local level and directing us to the dataset.

\hypertarget{references}{%
\section*{References}\label{references}}
\addcontentsline{toc}{section}{References}

\hypertarget{refs}{}
\leavevmode\hypertarget{ref-Anastassopoulou2020}{}%
Anastassopoulou, Cleo, Lucia Russo, Athanasios Tsakris, and Constantinos Siettos. 2020. ``Data-Based Analysis, Modelling and Forecasting of the Covid-19 Outbreak.'' \emph{PLOS ONE} 15 (3): 1--21. \url{https://doi.org/10.1371/journal.pone.0230405}.

\leavevmode\hypertarget{ref-Balcan21484}{}%
Balcan, Duygu, Vittoria Colizza, Bruno Gonçalves, Hao Hu, José J. Ramasco, and Alessandro Vespignani. 2009. ``Multiscale Mobility Networks and the Spatial Spreading of Infectious Diseases.'' \emph{Proceedings of the National Academy of Sciences} 106 (51): 21484--9. \url{https://doi.org/10.1073/pnas.0906910106}.

\leavevmode\hypertarget{ref-Balcan2010132}{}%
Balcan, Duygu, Bruno Gonçalves, Hao Hu, José J. Ramasco, Vittoria Colizza, and Alessandro Vespignani. 2010. ``Modeling the Spatial Spread of Infectious Diseases: The Global Epidemic and Mobility Computational Model.'' \emph{Journal of Computational Science} 1 (3): 132--45. \url{https://doi.org/https://doi.org/10.1016/j.jocs.2010.07.002}.

\leavevmode\hypertarget{ref-Benvenuto2020}{}%
Benvenuto, Domenico, Marta Giovanetti, Lazzaro Vassallo, Silvia Angeletti, and Massimo Ciccozzi. 2020. ``Application of the Arima Model on the Covid-2019 Epidemic Dataset.'' \emph{Data in Brief} 29: 105340. \url{https://doi.org/https://doi.org/10.1016/j.dib.2020.105340}.

\leavevmode\hypertarget{ref-box2015time}{}%
Box, George EP, Gwilym M Jenkins, Gregory C Reinsel, and Greta M Ljung. 2015. \emph{Time Series Analysis: Forecasting and Control}. John Wiley \& Sons.

\leavevmode\hypertarget{ref-Brodersen2015}{}%
Brodersen, Kay H., Fabian Gallusser, Jim Koehler, Nicolas Remy, and Steven L. Scott. 2015. ``Inferring Causal Impact Using Bayesian Structural Time-Series Models.'' \emph{Annals of Applied Statistics} 9: 247--74.

\leavevmode\hypertarget{ref-Chimmula2020}{}%
Chimmula, Vinay Kumar Reddy, and Lei Zhang. 2020. ``Time Series Forecasting of Covid-19 Transmission in Canada Using Lstm Networks.'' \emph{Chaos, Solitons \& Fractals} 135: 109864. \url{https://doi.org/https://doi.org/10.1016/j.chaos.2020.109864}.

\leavevmode\hypertarget{ref-durbin2012time}{}%
Durbin, James, and Siem Jan Koopman. 2012. \emph{Time Series Analysis by State Space Methods}. Vol. 38. Oxford University Press.

\leavevmode\hypertarget{ref-Fanelli2020}{}%
Fanelli, Duccio, and Francesco Piazza. 2020. ``Analysis and Forecast of Covid-19 Spreading in China, Italy and France.'' \emph{Chaos, Solitons \& Fractals} 134: 109761. \url{https://doi.org/https://doi.org/10.1016/j.chaos.2020.109761}.

\leavevmode\hypertarget{ref-Feroze2020}{}%
Feroze, Navid. 2020. ``Forecasting the Patterns of Covid-19 and Causal Impacts of Lockdown in Top Five Affected Countries Using Bayesian Structural Time Series Models.'' \emph{Chaos, Solitons \& Fractals} 140: 110196. \url{https://doi.org/https://doi.org/10.1016/j.chaos.2020.110196}.

\leavevmode\hypertarget{ref-Gupta2020}{}%
Gupta, Rajan, and Saibal Kumar Pal. 2020. ``Trend Analysis and Forecasting of Covid-19 Outbreak in India.'' \emph{medRxiv}.

\leavevmode\hypertarget{ref-Hamzah2020}{}%
Hamzah, FA Binti, C Lau, H Nazri, DV Ligot, G Lee, CL Tan, MKBM Shaib, et al. 2020. ``CoronaTracker: Worldwide Covid-19 Outbreak Data Analysis and Prediction.'' \emph{Bull World Health Organ} 1: 32. \url{https://doi.org/http://dx.doi.org/10.2471/BLT.20.255695}.

\leavevmode\hypertarget{ref-Yonar2020}{}%
Harun, Yonar, Yonar Aynur, Agah Tekindal Mustafa, and Tekindal Melike. 2020. ``Modeling and Forecasting for the Number of Cases of the Covid-19 Pandemic with the Curve Estimation Models, the Box-Jenkins and Exponential Smoothing Methods.'' \emph{EJMO} 4 (2): 160--65. \url{https://doi.org/10.14744/ejmo.2020.28273}.

\leavevmode\hypertarget{ref-hong2016probabilistic}{}%
Hong, Tao, Pierre Pinson, Shu Fan, Hamidreza Zareipour, Alberto Troccoli, and Rob J Hyndman. 2016. ``Probabilistic Energy Forecasting: Global Energy Forecasting Competition 2014 and Beyond.'' \emph{International Journal of Forecasting} 32: 896--913.

\leavevmode\hypertarget{ref-hyndman2020forecasting}{}%
Hyndman, Rob J, and George Athanasopoulos. 2020. \emph{Forecasting: Principles and Practice}. OTexts. \url{https://otexts.com/fpp3}.

\leavevmode\hypertarget{ref-hyndman2006another}{}%
Hyndman, Rob J, and Anne B Koehler. 2006. ``Another Look at Measures of Forecast Accuracy.'' \emph{International Journal of Forecasting} 22 (4): 679--88.

\leavevmode\hypertarget{ref-Ioannidis2020}{}%
Ioannidis, John P. A., Sally Cripps, and Martin A. Tanner. 2020. ``Forecasting for Covid-19 Has Failed.'' \emph{International Journal of Forecasting}. \url{https://doi.org/https://doi.org/10.1016/j.ijforecast.2020.08.004}.

\leavevmode\hypertarget{ref-Liu2020}{}%
Liu, Dianbo, Leonardo Clemente, Canelle Poirier, Xiyu Ding, Matteo Chinazzi, Jessica T Davis, Alessandro Vespignani, and Mauricio Santillana. 2020. ``A Machine Learning Methodology for Real-Time Forecasting of the 2019-2020 Covid-19 Outbreak Using Internet Searches, News Alerts, and Estimates from Mechanistic Models.'' \emph{arXiv Preprint arXiv:2004.04019}.

\leavevmode\hypertarget{ref-Martelloni2020}{}%
Martelloni, Gabriele, and Gianluca Martelloni. 2020. ``Modelling the Downhill of the Sars-Cov-2 in Italy and a Universal Forecast of the Epidemic in the World.'' \emph{Chaos, Solitons \& Fractals} 139: 110064. \url{https://doi.org/https://doi.org/10.1016/j.chaos.2020.110064}.

\leavevmode\hypertarget{ref-Massonnaud2020}{}%
Massonnaud, Clément, Jonathan Roux, and Pascal Crépey. 2020. ``COVID-19: Forecasting Short Term Hospital Needs in France.'' \emph{medRxiv}.

\leavevmode\hypertarget{ref-Moftakhar2020}{}%
Moftakhar, Leila, SEIF Mozhgan, and Marziyeh Sadat Safe. 2020. ``Exponentially Increasing Trend of Infected Patients with Covid-19 in Iran: A Comparison of Neural Network and Arima Forecasting Models.'' \emph{Iranian Journal of Public Health} 49: 92--100.

\leavevmode\hypertarget{ref-Mohd2020}{}%
Mohd, Mohd Hafiz, and Fatima Sulayman. 2020. ``Unravelling the Myths of R0 in Controlling the Dynamics of Covid-19 Outbreak: A Modelling Perspective.'' \emph{Chaos, Solitons \& Fractals} 138: 109943. \url{https://doi.org/https://doi.org/10.1016/j.chaos.2020.109943}.

\leavevmode\hypertarget{ref-Nabi2020}{}%
Nabi, Khondoker Nazmoon. 2020. ``Forecasting Covid-19 Pandemic: A Data-Driven Analysis.'' \emph{Chaos, Solitons \& Fractals} 139: 110046. \url{https://doi.org/https://doi.org/10.1016/j.chaos.2020.110046}.

\leavevmode\hypertarget{ref-fable2020}{}%
O'Hara-Wild, Mitchell, Rob Hyndman, Earo Wang, and Gabriel Caceres. 2020. \emph{Fable: Forecasting Models for Tidy Time Series}. \url{https://CRAN.R-project.org/package=fable}.

\leavevmode\hypertarget{ref-perc2020forecasting}{}%
Perc, Matjaž, Nina Gorišek Miksić, Mitja Slavinec, and Andraž Stožer. 2020. ``Forecasting Covid-19.'' \emph{Frontiers in Physics} 8: 127.

\leavevmode\hypertarget{ref-Makridakis2020}{}%
Petropoulos, Fotios, and Spyros Makridakis. 2020. ``Forecasting the Novel Coronavirus Covid-19.'' \emph{PLOS ONE} 15 (3): 1--8. \url{https://doi.org/10.1371/journal.pone.0231236}.

\leavevmode\hypertarget{ref-Salathe1002616}{}%
Salathe, Marcel, Linus Bengtsson, Todd J. Bodnar, Devon D. Brewer, John S. Brownstein, Caroline Buckee, Ellsworth M. Campbell, et al. 2012. ``Digital Epidemiology.'' \emph{PLOS Computational Biology} 8 (7): 1--3. \url{https://doi.org/10.1371/journal.pcbi.1002616}.

\leavevmode\hypertarget{ref-Sardar2020}{}%
Sardar, Tridip, Sk Shahid Nadim, Sourav Rana, and Joydev Chattopadhyay. 2020. ``Assessment of Lockdown Effect in Some States and Overall India: A Predictive Mathematical Study on Covid-19 Outbreak.'' \emph{Chaos, Solitons \& Fractals} 139: 110078. \url{https://doi.org/https://doi.org/10.1016/j.chaos.2020.110078}.

\leavevmode\hypertarget{ref-Sarkar2020}{}%
Sarkar, Kankan, Subhas Khajanchi, and Juan J. Nieto. 2020. ``Modeling and Forecasting the Covid-19 Pandemic in India.'' \emph{Chaos, Solitons \& Fractals} 139: 110049. \url{https://doi.org/https://doi.org/10.1016/j.chaos.2020.110049}.

\leavevmode\hypertarget{ref-Gleam2020}{}%
The Gleam Project. 2020. ``Covid-19 modelling, United States.'' \url{https://covid19.gleamproject.org/}.

\leavevmode\hypertarget{ref-Tomar2020}{}%
Tomar, Anuradha, and Neeraj Gupta. 2020. ``Prediction for the Spread of Covid-19 in India and Effectiveness of Preventive Measures.'' \emph{Science of the Total Environment} 728: 138762. \url{https://doi.org/https://doi.org/10.1016/j.scitotenv.2020.138762}.

\leavevmode\hypertarget{ref-Weissman2020}{}%
Weissman, Gary E, Andrew Crane-Droesch, Corey Chivers, ThaiBinh Luong, Asaf Hanish, Michael Z Levy, Jason Lubken, et al. 2020. ``Locally Informed Simulation to Predict Hospital Capacity Needs During the Covid-19 Pandemic.'' \emph{Annals of Internal Medicine}.

\leavevmode\hypertarget{ref-winkler1972decision}{}%
Winkler, Robert L. 1972. ``A Decision-Theoretic Approach to Interval Estimation.'' \emph{Journal of the American Statistical Association} 67 (337): 187--91.

\leavevmode\hypertarget{ref-WHO2020DashBoard}{}%
World Health Organisation. 2020. ``WHO Coronavirus Disease (COVID-19) Dashboard.'' \url{https://covid19.who.int/}.

\leavevmode\hypertarget{ref-Yousaf2020}{}%
Yousaf, Muhammad, Samiha Zahir, Muhammad Riaz, Sardar Muhammad Hussain, and Kamal Shah. 2020. ``Statistical Analysis of Forecasting Covid-19 for Upcoming Month in Pakistan.'' \emph{Chaos, Solitons \& Fractals} 138: 109926. \url{https://doi.org/https://doi.org/10.1016/j.chaos.2020.109926}.

\end{document}